# Single-mode lasers using parity-time-symmetric polarization eigenstates

Jean-François Bisson[*]

Yves Christian Nonguierma

Département de physique et d'astronomie, Faculté des sciences, Université de Moncton

18, av. Antonine-Maillet, Moncton, New Brunswick, Canada E1A 3E9

[*]Jean-francois.bisson@umoncton.ca

**ABSTRACT**

Anisotropic mirrors are used to form a laser resonator exhibiting non-Hermitian, parity-time (PT) symmetric, polarization states. The relative angle of the two mirrors' principal axes is used to control the degree of non-hermiticity. A sharp symmetry-breaking transition is observed at a specific angle, called the exceptional point, where the two states coalesce into a single polarization state and the interference pattern produced by counter-propagating (CP) waves vanishes. At a smaller angle, in the unbroken PT symmetry regime, the polarization state experiencing higher losses is suppressed. In the broken symmetry regime, the two polarization states coexist, but the orthogonality of the CP waves favors single longitudinal mode emission by suppressing the interference pattern of the standing wave. The two regimes meet at the exceptional point, where a unique polarization state exists in a resonator free from interference intensity pattern. Microchip PT-symmetric lasers operating at the exceptional point are thus an attractive solution to achieve single mode operation from a miniature monolithic device without any intra-cavity element.

# I. INTRODUCTION

Carl Bender and Stefan Boettcher showed in 1998 that parity-time (PT) reflection symmetric operators can exhibit, like Hermitian systems, entirely real eigenvalue spectrum, and therefore could not be ruled out as possible representations of observables in quantum mechanics [1]. PT-symmetric operators also exhibit spontaneous symmetry breaking when the value of some non hermiticity parameter is exceeded. It is in the field of optics that PT-symmetric systems have the most successfully been applied [see reviews in ref. [2-5] and references therein]. In laser science, PT-symmetry breaking was applied to generate single longitudinal mode laser operation in inhomogeneously broadened microring resonators [6,7]. In one instance, a PT-symmetric single mode laser was realized by carefully matching the gain of one micro-ring laser with the loss of the other coupled resonator such that one single mode experienced enhanced amplification from PT-symmetry breaking while the other competing modes were suppressed by remaining in the unbroken PT-symmetric region [6]. In a different development, single mode emission was achieved by delicately manipulating the gain and loss distribution into a whispering gallery mode laser utilizing the PT-symmetry breaking concept [7].

Here, we present an alternate route to achieve single mode laser operation that also makes use of PT-symmetry breaking but does not require a delicate manipulation of gain and loss between coupled resonators or their spatial distribution. We show that anisotropic laser mirrors can be used to form a laser resonator exhibiting non-Hermitian, PT symmetric, polarization states. By adjusting the relative angle of the two mirrors', the degree of non-hermiticity can be adjusted. In the unbroken PT-symmetry regime, dual polarization oscillation is suppressed, while in the broken PT-symmetry regime, single longitudinal mode operation is achieved by eliminating the axial intensity pattern of the standing wave. The two regimes meet at the transition point between both regimes, the exceptional point (EP), enabling single frequency operation.

In conventional linear laser cavity designs, the intensity pattern of the standing wave produces regularly spaced regions of undepleted inversion density that can be used by other axial modes to achieve laser oscillation despite their lower emission cross-section [8]. This generally gives rise to undesired multiple mode operation. Inserting quarter-wave plates in front of each laser mirror eliminates the contrast of the standing wave by making counter-propagating (CP) waves orthogonal [9]. This is the so-called *twisted-mode* operation, wherein the interference between CP left- or right-circularly polarized propagating eigen-waves produces a standing wave with axially uniform intensity and a linear polarization state that rotates like a twisted ribbon along the axial direction. But this scheme has the notorious drawback of not discriminating between the two coexisting eigen-polarization states. As a result, dual emission in both polarization states takes place; it is generally eliminated by placing a polarization-selective element between a λ/4 waveplate and a mirror [10-12]. Our concept gets rid of that competition between polarization states at the root by eliminating the very existence of dual polarization, while retaining the advantage of single longitudinal mode operation of the twisted-mode design.

The discovery of exceptional points (EP) of polarization goes back to the beginning of the 20$^{th}$ century in connection to the propagation of light in absorbing biaxial crystals [13]. This phenomenon, observed for some specific directions of light called *singular axes*, was correctly analyzed by Pancharatnam in 1955 [14]. In addition to demonstrating the existence of the coalescence of eigen-states of polarization in singular directions, Pancharatnam showed that light beams with polarization states other than the eigenstate gradually transform into the eigenvector as it propagates along a singular axis. Hence, Pancharatnam's discovery really appears to be a precursor to the recently proposed omnipolarizer [15]. The idea of PT-symmetry breaking in the polarization space is more recent and it generally aims at achieving active control of polarization [15-21], like compact polarization converters. It is generally realized with engineered metasurfaces or waveguides, wherein gain and loss are carefully balanced.

Our approach differs from previous works in that it does not require exquisite adjustment of gain and loss or complex nanofabrication steps. We also emphasize that, in contrast with previous single mode lasers based on PT-symmetry breaking, the selective symmetry breaking of one mode is not involved here. Indeed, in ref. [6], the resonators are designed such that only one mode strikes the right balance of gain and loss and undergoes selective symmetry breaking, thereby providing enhanced gain contrast and a mode selection mechanism. Here, the optical properties of the mirrors do not significantly change on the scale of the mode spacing, so several modes simultaneously experience the transition from unbroken to broken PT-symmetry at the exceptional point. The single mode selection is nevertheless enabled for homogeneously broadened active materials by the elimination of the axial spatial hole burning due to the orthogonality of the polarization states of the counterpropagating waves. The transition point between the two regions is identified as the privileged operation point where effective discrimination between dual polarization states and competing longitudinal modes is simultaneously achieved.

This paper is organized as follows. In section 2, we show that a resonator made of anisotropic mirrors can produce a characteristic polarization Jones matrix satisfying PT symmetry. First, we derive the round-trip Jones matrix of such a resonator and then we compare it with the general form of a PT-symmetric matrix for a two-by-two Jones matrix. This enables us to identify diattenuation and a $\pi$ phase shift between orthogonal principal axes of each mirror as the critical ingredients to achieve PT-symmetric eigenpolarization states. The relative orientation of the mirrors' principal axes, $\alpha$, is also identified as a flexible control parameter that enables to continuously span the transition between unbroken and broken PT-symmetry regions. In section 3, we describe the experimental conditions used to demonstrate such a laser system. We describe the laser resonator and the diagnostic method of the polarization state of the emitted beam and the experimental set-up used to determine the frequency spectrum of the emitted radiation in order to infer its mode content. In section 4, we show theoretical calculations and compare them with experimental results. First, we present the eigenvalues and corresponding polarization eigenstates inside the resonators. We relate the magnitude of the eigenvalues, which provide the round-trip losses of the resonator, with the measured threshold of laser oscillation. We also calculate the contrast of the standing wave inside the resonator and show that they become orthogonal in broken PT-symmetric region, thereby providing efficient elimination of spatial hole burning. We measure the eigen-polarization states of the emitted beam and experimentally confirm the suppression of one polarization state with higher losses in the unbroken PT-symmetric region and the coalescence of eigenvectors at the exceptional point. More importantly, we demonstrate the possibility of single mode laser operation exploiting PT-symmetry breaking. We present the emission spectrum of the emitted radiation and confirm the prediction that the multi-longitudinal mode emission is suppressed in the broken PT-symmetric region. We argue that the exceptional point is a privileged point where both dual-polarization and multimode emission are prevented. We conclude in section 5 with some discussion about the limitations of our concept. We discuss the anticipated benefit of reducing the length of the resonator by using nanostructured anisotropic thin films as laser mirrors, the effect of small errors in the phase shift of the mirrors and that of thermal birefringence in the active medium on the performance of the device.

## II. THE PT-SYMMETRIC, TWISTED-MODE LASER RESONATOR

We consider a standing wave laser resonator made of two linearly anisotropic mirrors. The optical response of a non-depolarizing laser mirror can be modelled by a two-by-two Jones matrix; they are represented by diagonal Jones matrices in their principal basis as:

$$M_1 = \begin{pmatrix} r_{11} & 0 \\ 0 & r_{12} \end{pmatrix}_{xy} \quad (1)$$

and

$$M_2 = \begin{pmatrix} r_{21} & 0 \\ 0 & r_{22} \end{pmatrix}_{xy}, \quad (2)$$

where each coefficient is complex. One mirror is rotated with respect to the other around the resonator's optical axis by some angle $\alpha$ and we calculate the Jones matrix of an intra-cavity round-trip, assuming there are no polarizing element inside the resonator (thermal birefringence inside the active material is assumed negligible). In our convention, an isotropic mirror is noted as : $\begin{pmatrix} 1 & 0 \\ 0 & -1 \end{pmatrix}_{xy \to x'y'}$, the minus sign arising from the fact that the polarization state at reflection is expressed in new coordinates $(x',y',z')$ where the y and $z$ axes are reversed at the reflection. In the left-right circular basis, we obtain:

$$M_1 = \frac{1}{2}\begin{pmatrix} 1 & -i \\ 1 & i \end{pmatrix}\begin{pmatrix} r_{11} & 0 \\ 0 & r_{12} \end{pmatrix}\begin{pmatrix} 1 & 1 \\ i & -i \end{pmatrix} = \frac{1}{2}\begin{pmatrix} r_{11}+r_{12} & r_{11}-r_{12} \\ r_{11}-r_{12} & r_{11}+r_{12} \end{pmatrix}_{lr} \quad (3a)$$

and

$$M_2 = \frac{1}{2}\begin{pmatrix} r_{21}+r_{22} & r_{21}-r_{22} \\ r_{21}-r_{22} & r_{21}+r_{22} \end{pmatrix}_{lr}. \quad (3b)$$

If one mirror is rotated by angle α, with respect to the other, then the Jones matrix for a round trip, $J$, is given by [22]:

$$J = TM_2 TTM_1 T, \quad (4)$$

where $T$ is the rotation matrix by angle α/2, given by:

$$T(\alpha/2) = \begin{pmatrix} \exp(i\alpha/2) & 0 \\ 0 & \exp(-i\alpha/2) \end{pmatrix} \quad (5)$$

The propagation inside the resonator in free space or inside a homogeneous active material corresponds to a multiple of the identity matrix and does not play any role apart from a constant phase factor and is thus ignored. The computation of (4) with (3) and (5) gives:

$$J = \frac{1}{4}\begin{pmatrix} (r_{21}+r_{22})(r_{11}+r_{12})\exp(2i\alpha)+(r_{21}-r_{22})(r_{11}-r_{12}) & (r_{21}+r_{22})(r_{11}-r_{12})\exp(i\alpha)+(r_{21}-r_{22})(r_{11}+r_{12})\exp(-i\alpha) \\ (r_{21}+r_{22})(r_{11}-r_{12})\exp(-i\alpha)+(r_{21}-r_{22})(r_{11}+r_{12})\exp(i\alpha) & (r_{21}+r_{22})(r_{11}+r_{12})\exp(-2i\alpha)+(r_{21}-r_{22})(r_{11}-r_{12}) \end{pmatrix}_{lr}.$$

(6)

Now, it is interesting to ask the conditions required for $J$ to be PT-symmetric. If we have the general form of a PT-symmetric Jones matrix and compare it to Eq. (6), we might be able to specify the optical properties of the mirrors (i.e., the $r_{ij}$ values) and we might find an experimentally accessible control parameter that makes it possible to continuously cover the unbroken and broken PT-symmetric regions. We shall see that such a control parameter does exist and is closely linked to the torsion angle, α.

We define a matrix $J$ as PT-symmetric if it satisfies the commutation relation [1]:

$$(PT)J - J(PT) = 0. \quad (7)$$

where *P* is the parity operator and T is the time-reversal operator, defined here as the complex conjugate. Mostafazadeh [23], and Wang [24] derive the general form for a PT-symmetric two-by-two matrix as :

$$J_{PT} = \begin{pmatrix} A + B\cos\theta - iC\sin\theta & (B\sin\theta + iC\cos\theta + iD)\exp(-i\varphi) \\ (B\sin\theta + iC\cos\theta - iD)\exp(i\varphi) & A - B\cos\theta + iC\sin\theta \end{pmatrix}, \quad (8)$$

where *A*, *B*, *C* and *D* can take any real value, $0 \leq \theta < \pi$ and $0 \leq \varphi < 2\pi$. Wang used a different definition of the time-reversal operator. We present another derivation in Appendix A, consistent with our definition of the time-reversal operator, which is based solely on the invariance of the eigenvalues by unitary transformations of polarization states.

By comparing eqs. (6) and (8), we find that the round-trip operator $J_{RT}$ can be made PT-symmetric by taking all $r_{ij}$ real. The comparison gives:

$$A = \frac{1}{4}\left[(r_{21} + r_{22})(r_{11} + r_{12})\cos(2\alpha) + (r_{21} - r_{22})(r_{11} - r_{12})\right], \tag{9a}$$

$$B = \frac{r_{21}r_{11} - r_{22}r_{12}}{2}\cos\alpha, \tag{9b}$$

$$C = -\frac{1}{4}\left[(r_{21} + r_{22})(r_{11} + r_{12})\sin(2\alpha)\right], \tag{9c}$$

$$D = \frac{r_{22}r_{11} - r_{21}r_{12}}{2}\sin\alpha, \tag{9d}$$

$$\theta = \pi/2, \tag{9e}$$

$$\varphi = 0. \tag{9f}$$

We note that *J* is Hermitian when *C*=0; hence, *C* determines the degree non-hermiticity. We may define an order parameter $\chi$ as [Cf. Appendix A]:

$$\chi \equiv C^2/(B^2 + D^2). \tag{10}$$

The condition for unbroken symmetry, where *J* and *PT* operators share the same eigenvectors and the eigenvalues are real, is given by:

$$\chi \leq 1. \tag{11}$$

It is instructive to compute χ for a simple case of a pair of identical mirrors, i.e., $r_{21} = r_{11} \equiv r_1$ and $r_{22} = r_{12} \equiv r_2$ so that $M_1 = M_2 = \begin{pmatrix} r_1 & 0 \\ 0 & r_2 \end{pmatrix}_{xy}$. From Eqs. (9-10), we then find:

$$\chi = \frac{(r_1 + r_2)^2 \sin^2(\alpha)}{(r_1 - r_2)^2}. \tag{12}$$

One can see that the non hermiticity is controlled by the torsion angle, α, and the dichroism, $|r_1-r_2|$. The transition occurs at the exceptional point, $α=α_{EP}$, where χ=1:

$$\alpha_{EP} = \pm \arcsin\left(\frac{r_2 - r_1}{r_1 + r_2}\right). \tag{13}$$

Note that the transition between unbroken and broken PT symmetry at the exceptional points exists only if :

1. $r_1$ and $r_2$ have the same sign, which, in our convention, implies that a π phase shift exists between orthogonal axes;

and

2. $r_1 \neq r_2$, i.e., diattenuation exists.

Resonators made of mirrors without phase shift, where $r_1$ and $r_2$ take real values of opposite signs, are not interesting despite their PT-symmetric character, because the PT-symmetry cannot be broken for any α value, since χ<1 for any α. Likewise, resonators without diattenuation, where $r_1=r_2$, are not suitable for controlling PT-symmetry, since χ>1, for any α value, except for the trivial case, α=0 and π/2, where the resonator is isotropic.

### III. EXPERIMENTAL METHODS

The laser resonator is made of a flat rear mirror that is transparent to pump light (λ=933nm) and highly reflective at laser wavelength (λ=1030nm) and a concave output coupler with a 100-mm radius of curvature and a 92% reflectance at the laser wavelength. An antireflection-coated zero-order quarter-wave plate is placed in front of each mirror to create a π reflection phase shift between orthogonal axes. A 1-mm thick glass plate inclined at 60º near the Brewster angle with respect to normal incidence is placed immediately in front of the output-coupler-λ/4 plate combination: the three component together simulate a deattenuating and birefringent mirror with π phase shift with reflection matrix:

$$M_1 = \begin{pmatrix} 0.70 & 0 \\ 0 & 1.0 \end{pmatrix}, \tag{14}$$

in the horizontal-vertical basis, while the rear mirror-λ/4 combination simulates a birefringent mirror with π phase shift of the form:

$$M_2 = \begin{pmatrix} 1.0 & 0 \\ 0 & 1.0 \end{pmatrix}, \tag{15}$$

i.e, without any dichroism, in the basis of its principal axes. The rear mirror is mounted on a rotation stage in order to control the relative orientation α of the two mirrors and study its effects on the laser characteristics. A 1-mm-thick, antireflection-coated, 10 at. % $Yb^{3+}$-doped $Y_3Al_5O_{12}$ ceramic is placed between the two mirrors on top of the λ/4 waveplate of the rear mirror. The total resonator length is $L \approx 2.5$ cm. The knowledge of the FSR enabled us to assign regularly interference rings to The mode size of the $TEM_{00}$ Gaussian mode is estimated from the resonator geometry alone to be about $w_0=120$μm near the rear mirror. Light emitted from a fiber-coupled laser diode emitting at 933 nm is concentrated on the active material in an end-pumped scheme using a pair of plano-convex lenses to match the size of the $TEM_{00}$ fundamental mode inside the active medium. In order to minimize the generation of heat and thermal birefringence inside the active material, the pump is turned on during 10 μs for most experiments and this is repeated every 125 μs (8% duty cycle). Sometimes the pump pulse duration needed to be increased or decreased in order to make one of the two eigen-polarization modes to come out and lend itself to ellipsometric analysis. Part of the emitted laser radiation is sent onto a silicon-based photodetector to enable the detection of the onset of the laser oscillation. The minimum driving current of the laser diode required to obtain the laser oscillation is sought by optimizing the alignment of the cavity mirrors for every set α value. This, combined with the knowledge of the current-power characteristics and the fraction of absorbed pump power by the active element enable us to plot the threshold absorbed pump power as a function of angle α.

For the determination of the polarization state outside the resonator, the emitted radiation is transmitted through a quarter wave plate, called a compensator, followed by a polarizer, called an analyzer, each mounted on a rotation stage that allowed us to adjust the rotation angle in order to reach as close an extinction of the transmitted beam as possible. Extinction can be obtained first by converting the generally elliptical beam into a rectilinear polarization by aligning the compensator's fast axis with one axis of the elliptical pattern of the transverse electric field vector and then by seeking extinction by rotating the analyzer. Then, the angle of the analyzer at extinction is subtracted from the angle of the fast axis of the compensator (ξ) to produce the angle ψ. Then, the *x*, *y*, *z* coordinates on the Poincaré coordinates are determined by using [25]:

$$\begin{pmatrix} x \\ y \\ z \end{pmatrix} = \begin{pmatrix} \cos(2\xi)\cos(2\psi) \\ \sin(2\xi)\cos(2\psi) \\ -\sin(2\psi) \end{pmatrix}. \tag{16}$$

We found that the two states in the broken PT-symmetric region coexisted and randomly hopped from one to the other; although a detrimental effect, this coexistence enabled us to measure ξ and χ separately for each of them.

The calculated external eigenstate vector is obtained by multiplying the internal eigenstate vector incident on the output coupler by matrix:

$$T_{M_1} = \begin{pmatrix} \sqrt{0.70} & 0 \\ 0 & -i \end{pmatrix}, \tag{17}$$

where the upper left term accounts for diattenuation and the second for the quarter wave phase advance in the *y* direction.

The polarization states of counter-propagating (CP) waves of each mode is also an important parameter for laser operation, because the proximity of their polarization state determines the axial intensity contrast of the standing wave pattern and the possibility of multiple longitudinal mode operation. If we assume that the intensity of the counter propagating waves is nearly equal, the visibility $V$ of the interference pattern is equal to the magnitude of the Hermitian scalar product of the CP waves of each mode $i$ [22]:

$$V \equiv \frac{I_{max} - I_{min}}{I_{max} + I_{min}} \cong \left\| u_{i+}^\dagger u_{i-} \right\|, \tag{18}$$

where $u_{i+}$ and $u_{i-}$ denote the Jones vector of the mode $i$ propagating in the positive and negative $z$ directions and the symbol † denotes the conjugate transpose.

For the determination of the emission spectrum, the experimental set-up and the calculation used to convert the fringe position in the camera into a frequency shift are explained in detail in ref. [26]. In summary, in order to analyze the emission spectrum, the unabsorbed pump light at 933 nm is first eliminated by using a low-pass filter cutting light at λ<950nm. The emitted beam is concentrated onto a Fabry-Perot (FP) étalon (free spectral range $FSR=$ 30 GHz, finesse $F=30$ at 1030 nm) with a microscope objective with numerical aperture NA=0.2. A set of sharp circular interference fringes corresponding to the matching of the resonance condition of the FP etalon can be observed on a CCD camera placed at the focal plane of a $f=70$-mm lens. In the broken PT-symmetry region, where competing polarization states could be observed, the dual polarization operation appeared as a splitting in the frequency emission; then, the compensator and the analyzer are adjusted to select only one of the two eigenmodes. The timing of the trigger of the capture is adjusted such that the chosen mode was emitted during the capture window, which generally last 20 μs. At the exceptional point, the two polarization states merge together, while inside the unbroken symmetry region, only one state can oscillate, the other one being suppressed by the difference of intra-cavity loss between them.

## IV. RESULTS

The calculated magnitude and phase of the eigenvalues are shown in Fig. 1a as a function of α for our experimental parameters. In the unbroken PT-symmetry region, $|\alpha|<\alpha_{EP}$, eigenvalues are pure real numbers; one polarization state suffers higher losses than the other and is expected to be suppressed in laser operation when the saturation of the active medium takes place. In the broken PT-symmetry region, $|\alpha|>\alpha_{EP}$, eigenvalues are complex conjugates, which suggests that dual polarization emission will take place. The magnitude of the eigenvalue is smaller than that of the preferred polarization state in the unbroken PT symmetry region, which implies that a lower threshold of laser oscillation should take place in the latter.

This is indeed what is experimentally found when measuring the threshold pump power for laser oscillation, as a function of α, Fig. 1b. There are two regions of unbroken symmetry corresponding to the fast axes of the quarter waveplates parallel or orthogonal. The oscillation threshold is found to be constant in the broken PT-symmetry region and sharply drops inside the PT-symmetric region, in accordance with the magnitude of the calculated eigenvalues.

The calculated eigenvectors of the two modes inside the resonators are shown in Fig. 2a as ($x,y,z$) coordinates on the Poincaré sphere. The polarization states remain rectilinear ($z=0$) in the unbroken PT-symmetry region and their planes of polarization rotate towards each other as α increases; then, they merge together at the EP to form a single, degenerate diagonal polarization state ($x=0$, $y=1$, $z=0$); afterwards, they split again in the broken symmetry region to approach circular left and right ($z=\pm1$) at larger α values.

We analyzed the polarization states of the emitted beam by seeking extinction with a compensator and an analyzer. As expected, only one polarization state of the emitted beam was observed in the PT-symmetric region because of the gain saturation, while the emission randomly hopped between both eigen-polarization states in the broken PT-symmetry region, allowing for the characterization of each polarization state. The experimental data, shown in Fig. 2b, are in good agreement with the theory and clearly show the coalescence of both states at the exceptional point located near $\alpha=\pm5°$.

The visibility of the interference pattern for each mode, as calculated from the magnitude of the Hermitian scalar product of the CP waves of each mode, is shown in Fig. 3 as a function of *α*. The corresponding trajectories of the eigenvectors of the counterpropagating waves of one mode are shown on the Poincaré sphere in the inset of

Fig. 3. The CP waves of each eigenmode, initially parallel at $\alpha=0$, become increasingly orthogonal as $|\alpha|$ increases; at the EP, the CP waves become perfectly orthogonal and remain so throughout the broken PT-symmetry region. The significance of this result is that spatial hole burning can be suppressed in the broken PT-symmetry region [8-12,22,26,27]. Therefore, the EP appears to be a privileged operating condition where single longitudinal emission in a single polarization state can be achieved.

The emission spectrum at different α values using a high finesse Fabry-Perot etalon is shown in Fig. 4. When multi-longitudinal mode operation took place, the frequency interval was equal to the FSR (≈6GHz), or to integral number of FRS. The transition between the multimode emission in the unbroken PT-symmetric region to nearly single longitudinal mode operation in the broken-symmetry region is clearly visible. In the broken symmetry region ($\alpha = -15°$), each polarization state can be detected separately with our polarization analyzer but each polarization state was almost single mode. At α=0°, the emission spectrum is highly multimode but only one polarization state could be observed. Near the exceptional point (α ≈ 5°), the emission only shows one polarization state and only two longitudinal modes could be detected.

## V. DISCUSSION AND CONCLUDING REMARKS

In summary, we have shown theoretically and experimentally that PT-symmetry breaking of the polarization eigenstates can be achieved with a twisted laser resonator made of anisotropic mirrors having both a π phase shift between principal axes of each mirrors and diattenuation. The torsion angle between the two mirrors is a versatile control parameter that enables one to probe the transition between unbroken and broken PT-symmetry near the exceptional point and control the properties of laser emission. Dual polarization oscillation is suppressed in the unbroken PT-symmetry region, while multiple longitudinal mode emission is suppressed in the broken PT-symmetry region. Single mode laser operation can be achieved at the exceptional point by suppressing dual polarization emission and axial spatial hole burning.

However, we experimentally found that the transition from multimode to single longitudinal mode operation is not as sharp as one would expect. Three factors may explain this. First, the presence of intra-cavity elements makes the resonator rather long (*L*=2.5 cm, for a free spectral range of 6 GHz). As a result, the tiny frequency spacing between neighbouring modes promotes their competition, which materializes as mode hopping by thermal instabilities. The second factor is the imperfections of the optical elements, especially the λ/4 waveplates. The third factor is thermally-induced birefringence inside the active medium.

Earlier results with twisted-mode lasers indicate that using millimeter-long microchip lasers will eliminate both issues [27]. This will require the use of anisotropic thin film mirrors or nanofabrication techniques in order to get rid of the intracavity elements we are currently using. Existing technologies such as Glancing Angle Deposition [28] or diffractive optical elements etched into a dielectric multilayer such as circular gratings [29], resonant gratings [30] or photonic crystals [31-32] could be harnessed to realize such mirrors.

We investigated by numerical simulations the effect of a small phase shift $\varphi$ between diagonal entries of the Jones matrix:

$$\begin{pmatrix} |r_{11}| & 0 \\ 0 & |r_{12}|\exp(i\varphi) \end{pmatrix}_{xy}. \tag{19}$$

It turns out that the PT-symmetric behavior can adversely be affected by a small $\varphi$ value. This is illustrated in Fig. 5a and 5b for a phase shift of $2\pi/300$, corresponding to the specifications (λ/4±λ/300) of our quarter wave plates, and otherwise for our experimental conditions. The transition at the exceptional point is smoothened, the degeneracy of the eigenvalues and eigenvectors is lifted, and the counterpropagating waves of each mode are not perfectly orthogonal beyond the EP anymore. Hence, an error on the required π phase shift between orthogonal axes of only $2\pi/300$ is sufficient to lift the degeneracy at the EP and to deteriorate the uniformity of the standing wave. However, we also noted that using mirrors having opposite phase shifts, such as:

$$\begin{pmatrix} |r_{11}| & 0 \\ 0 & |r_{12}|\exp(i\varphi) \end{pmatrix}_{xy} \tag{20}$$

and

$$\begin{pmatrix} |r_{21}|\exp(i\varphi) & 0 \\ 0 & |r_{22}| \end{pmatrix}_{xy}, \tag{21}$$

has almost no detrimental effect on the eigenvalues, eigenvectors and the contrast of the standing wave pattern, and are therefore a good approximation of PT-symmetric matrices, Cf. Fig. 5. Using a shorter resonator will also relax tolerances on the mirrors' optical properties because the requirement for a uniform axial intensity pattern is not necessary to achieve single mode operation, provided the gain difference between neighboring modes is sufficient [22].

Thermally-induced birefringence was also ignored in our calculations and it is worth asking whether this was justified. This phenomenon arises from the heat deposited into the active material by the absorbed pump light. The inhomogeneous temperature profile produced by heat diffusion induces thermal strain inside the active material, which in turn, causes a spatially inhomogeneous thermally-induced birefringence by the photo-elastic effect [33]. This modifies the state of polarization of light passing through the active medium in an inhomogeneous manner. We did witness some depolarization of the emitted beam by measuring the extinction ratio of each eigenmodes with our polarization analyzer. We expected the phenomenon to worsen at higher pump power, but the opposite trend was observed. In the broken PT-symmetry region, the polarization extinction ratio (PER) was measured to be around 100 just above the oscillation threshold and improved steadily at higher pump power to reach 200 at a pump power of 1.5 times the oscillation threshold. In the unbroken PT-symmetric region, the measured PER value near α=0° was in the order of 1000 just above the oscillation threshold and increased to more than 3000 at three times the threshold pump power. That the PER was much higher in the unbroken region can be explained by the observation, made by Clarkson [34], that placing a λ/4 plate on one side of the resonator and a polarizer aligned with one axis of the λ/4 waveplate on the other side reduces depolarization losses by orders of magnitude. The mechanism is as follows:

Depolarization losses for horizontally polarized incident light are generally zero at 0° and 90° azimuthal locations around the pump axis because these positions have their principal axes of the thermally-induced birefringence aligned with the horizontal and vertical directions; conversely, the depolarization losses are usually highest in the diagonal azimuthal locations (i.e. ±45° with respect to the polarizer axes). However, at those azimuthal locations, the incoming vertically polarized beam splits into equal amount of diagonal components that undergo different amounts of phase shift, but these components are exchanged by rotation of the plane of polarization by 90° when traveling back and forth through the λ/4 and the phase shifts are also exchanged when passing through the active medium in the return trip, resulting into negligible depolarization losses at ±45° and ±135° as well. This scenario takes place here too near $α=0$ because the eigenvector at 0° is vertically polarized due to the presence of the Brewster plate, which acts as a polarizer; as α increases, the polarization states rotate (Fig. 2) and this scheme becomes not as effective; this explains why the depolarization losses are higher in the unbroken region. Why the PER improves at higher pump power is not understood at this time; nevertheless, the fact that the PER was high, combined with our experimental findings of a sharp transition at α≈±5°, and excellent agreement of the measured and the calculated polarization states for any α value support both the neglect of depolarization by thermal birefringence and it supports our model of eigenpolarization state based on our linear model.

## APPENDIX A: THE PT-SYMMETRIC JONES MATRIX AND THE CONTROL PARAMETER OF NON-HERMITICITY.

We present a derivation of the general PT-symmetric matrix (Eq. 8) and the control parameter χ (Eq. 10). We define a matrix $J$ as PT-symmetric if it satisfies the commutation relation [1]:

$$(PT)J - J(PT) = 0. \tag{A1}$$

where $P$ is the parity operator and T is the time-reversal operator, defined here as taking the complex conjugate, $^*$. Eq. (A1) is equivalent to:

$$PJ^* = JP. \tag{A2}$$

Now, we want to derive a general form of a PT-symmetric Jones matrix $J$ that can be expressed in any orthogonal basis of Jones vectors. First, we try a particular form of the $P$ matrix:

$$P = \begin{pmatrix} 1 & 0 \\ 0 & -1 \end{pmatrix}. \tag{A3}$$

This choice is justified by the generally accepted properties of a parity operator: it is Hermitian, unitary and it is an involution (i.e., it is its own inverse). If we write :

$$J = \begin{pmatrix} a & b \\ c & d \end{pmatrix}, \tag{A4}$$

where $a,b,c$ and $d$ coefficients are complex in general; then condition of PT-symmetry, eq. (A2), implies that $a$ and $d$ be real, and $b$ and $c$ be imaginary. Hence, we write:

$$J = \begin{pmatrix} \eta & i\beta \\ i\delta & \gamma \end{pmatrix}, \tag{A5}$$

where $\eta$, $\beta$, $\gamma$ and $\delta$ are all real. Now, one is free to express $J$ in any basis of one's choice. An orthogonal basis of Jones vectors can be parametrized as:

$$u_1 = \begin{pmatrix} \cos\theta/2 \\ \sin\theta/2 \exp(i\varphi) \end{pmatrix} \tag{A6a}$$

and

$$u_2 = \begin{pmatrix} -\sin\theta/2 \\ \cos\theta/2 \exp(i\varphi) \end{pmatrix}, \tag{A6b}$$

where angles $\theta$ ($0 \le \theta < \pi$) and $\varphi$ ($0 \le \varphi < 2\pi$) uniquely determine any polarization state. Hence, the corresponding unitary transformation is :

$$R = \begin{pmatrix} \cos\theta/2 & -\sin\theta/2 \\ \sin(\theta/2)\exp(i\varphi) & \cos\theta/2\exp(i\varphi) \end{pmatrix} \tag{A7}$$

we have :

$$J'(\theta,\varphi) = RJR^{-1} = \begin{pmatrix} \cos\theta/2 & -\sin\theta/2 \\ \sin(\theta/2)\exp(i\varphi) & \cos\theta/2\exp(i\varphi) \end{pmatrix} \begin{pmatrix} \eta & i\beta \\ i\delta & \gamma \end{pmatrix} \begin{pmatrix} \cos\theta/2 & \sin(\theta/2)\exp(-i\varphi) \\ -\sin\theta/2 & \cos\theta/2\exp(-i\varphi) \end{pmatrix}$$
. (A8)

A cumbersome but straightforward calculation of the eq. (A8) gives :

$$J = \begin{pmatrix} A + B\cos\theta - iC\sin\theta & (B\sin\theta + iC\cos\theta + iD)\exp(-i\varphi) \\ (B\sin\theta + iC\cos\theta - iD)\exp(i\varphi) & A - B\cos\theta + iC\sin\theta \end{pmatrix}, \tag{A9}$$

where :

$$A = \frac{\eta + \gamma}{2}, \tag{A10a}$$

$$B = \frac{\eta - \gamma}{2}, \tag{A10b}$$

$$C = \frac{\beta + \delta}{2} \tag{A10c}$$

and

$$D = \frac{\beta - \delta}{2}. \tag{A10d}$$

Note that $A$, $B$, $C$, and $D$ can take any real value. Note that the $P$ matrix also changes form when using a different basis. Let $P'$ be the $P$ matrix in the new basis. The commutation operation eq. (A2) implies that:

$$PR^{-1*}J^*R^* = R^{-1}JRP, \tag{A11a}$$

which in turn implies :

$$RPR^{-1*}J'^* = J'RPR^{-1*}. \tag{A11b}$$

By comparing eq. (A2) with eq. (A11b), we find:

$$P' = RPR^{-1*}. \tag{A12}$$

With the chosen unitary operation $R$, $P'$ becomes:

$$P' = RPR^{-1*} = \begin{pmatrix} \cos\theta & \sin\theta \exp(i\varphi) \\ \sin\theta \exp(i\varphi) & -\cos\theta \exp(2i\varphi) \end{pmatrix}. \tag{A13}$$

It is noteworthy that the general form of $P$ is neither unitary, nor Hermitian, nor an involution. Now, the unbroken PT symmetry corresponds to the conditions for the PT operator and $J$ to share a common set of eigenvectors. In order to identify the constraints on the $A$, $B$, $C$, $D$ parameters for the unbroken symmetry to be valid, it is convenient to choose a specific unitary transformation: that is with $\theta=0$ and $\varphi=\pi/2$ in eqs. (A9). One obtains:

$$J' = \begin{pmatrix} A+B & C+D \\ -C+D & A-B \end{pmatrix}, \tag{A14}$$

i.e., the four entries are real without further restrictions and the corresponding $P$ operator is:

$$P' = \begin{pmatrix} 1 & 0 \\ 0 & 1 \end{pmatrix}, \tag{A15}$$

which is the identity matrix. The eigen-vectors of the $P'T$ operator are then just those of $T$, the complex conjugate operator. Hence, the entries of the $P'T$ eigenvectors are any pair of coefficients with the same phase, or equivalently, any pair of real entries with a common global phase factor. Now, it is easy to show that the condition for the eigenvectors to $J'$ to also have pure real eigenvectors is that their eigenvalues be real. This condition is ensured by forcing the discriminant of the characteristic equation:

$$\begin{vmatrix} A+B-\lambda & C+D \\ -C+D & A-B-\lambda \end{vmatrix} = 0 \Leftrightarrow \lambda^2 - 2A\lambda + A^2 - B^2 + C^2 - D^2 = 0 \tag{A16}$$

to be positive, i.e.,

$$B^2 - C^2 + D^2 > 0. \tag{A17}$$

Defining the $\chi$ parameter as:

$$\chi \equiv C^2 / (B^2 + D^2) \tag{A18}$$

from which the condition for real eigenvalues:

$$\chi \leq 1 \tag{A19}$$

directly arises. This is the general condition for the unbroken PT symmetry, since the eigenvalues are invariant by a similarity transformation. Matrix $J$ becomes defective when $C^2 = B^2 + D^2$ and this corresponds to the exceptional point (EP) discussed in the main text. One could argue that, with the definition (A13), PT is hardly a parity-time reflection symmetry operator. However, the important thing here is the condition of the existence of pure real eigenvalues and the spontaneous breaking of ''PT'' symmetry at some value of the control parameter ($\chi = 1$).

**ACKKNOWLEDGMENT**

The authors acknowledge financial support from the Natural Sciences and Engineering Research Council of Canada (Grant 5215).

**FIGURES AND CAPTIONS**

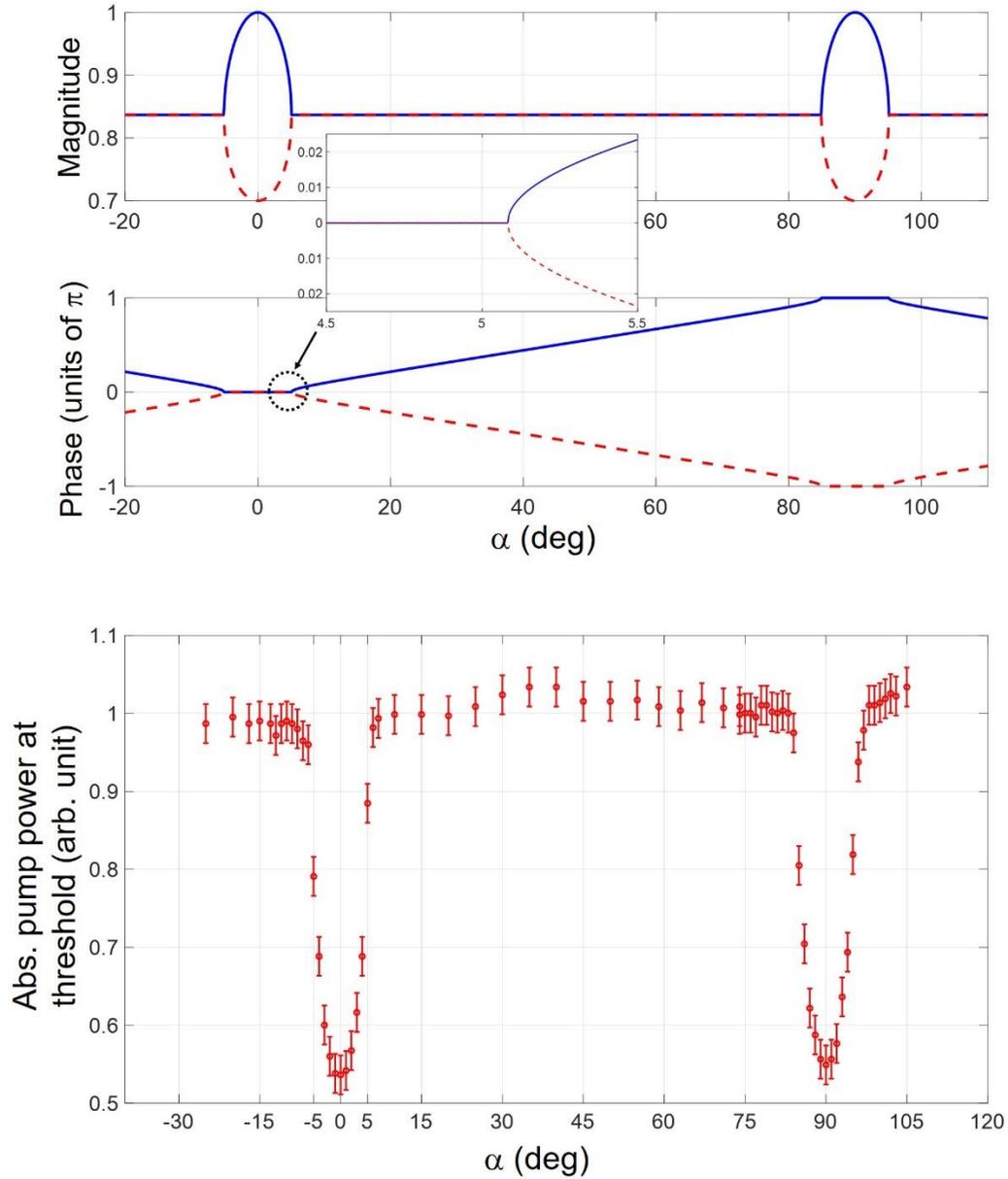

Fig. 1. (a) Calculated magnitude (top) and phase (bottom) of the eigenvalues of the Jones matrix of the round-trip. The square-root dependence of the phase in the neighbourhood of the exceptional point is shown in the inset. (b) Experimental threshold of laser oscillation as a function of α. A sharp drop in threshold is noticed at $α=α_{EP}≈±5°$ and also when one mirror is rotated by 90° in agreement with the larger magnitude of the eigenvalues in the region of unbroken PT-symmetry.

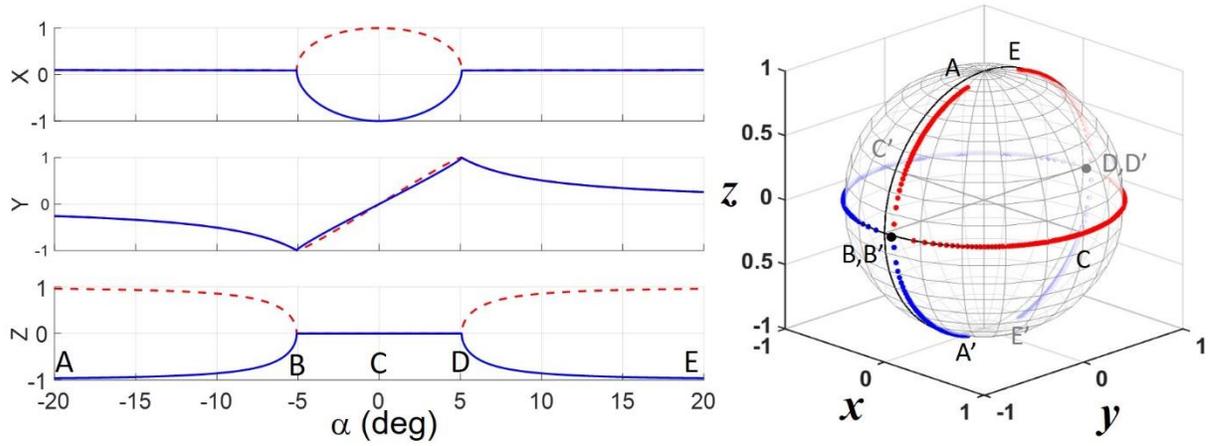

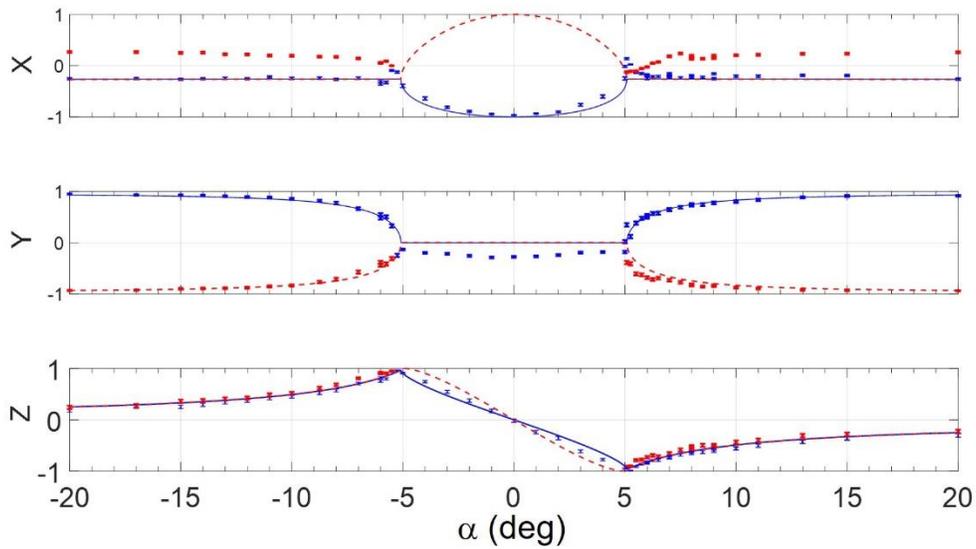

Fig. 2. (a) Calculated x,y,z coordinates on the Poincaré sphere of the two eigen polarization states (solid blue and dashed red) inside the resonator for the waves heading towards the output coupler. The two eigenstates merge at the exceptional point near ±5° (points B and D). The corresponding trajectory of the modes on the Poincaré sphere is shown on the right.

(b) Experimental (dots) and calculated (lines) x,y,z coordinates on the Poincaré sphere of the two eigen polarization states (solid and dashed lines) of the emitted radiation outside the resonator. The two eigenstates merge at the exceptional point near ±5° where the polarization state is circular.

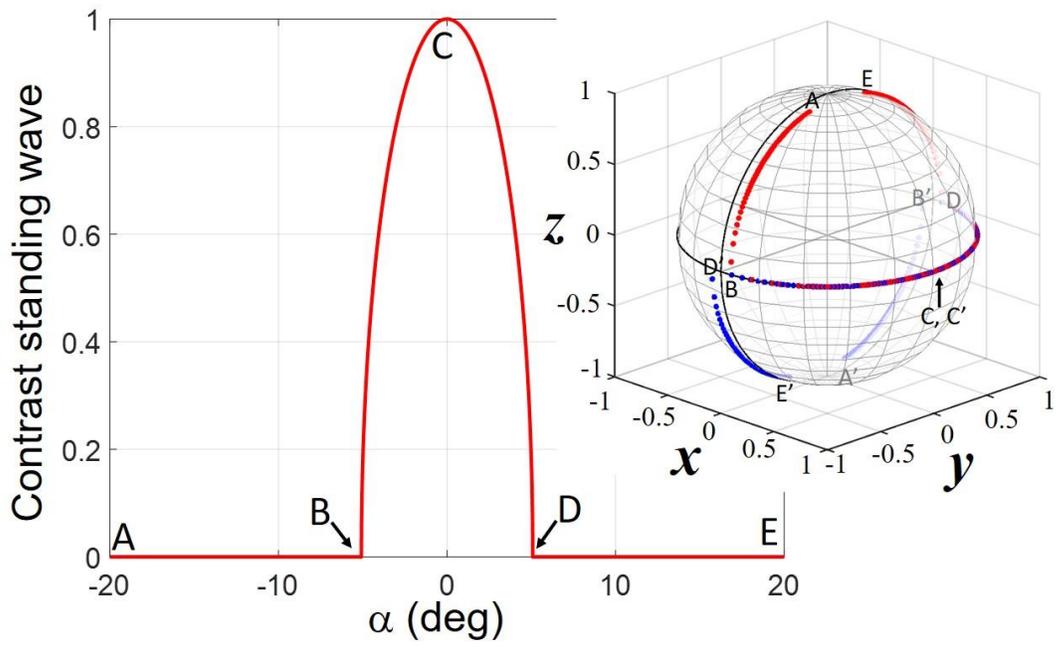

Fig. 3. Contrast of the standing wave, identical for each eigenmodes, as a function of the angle α. The counter-propagating waves become orthogonal at $α=α_{EP}$ and remain so at larger α values. The corresponding trajectories of the counter-propagating waves (red: the wave going towards the output coupler) for mode 1 on the Poincaré sphere is shown on the right.

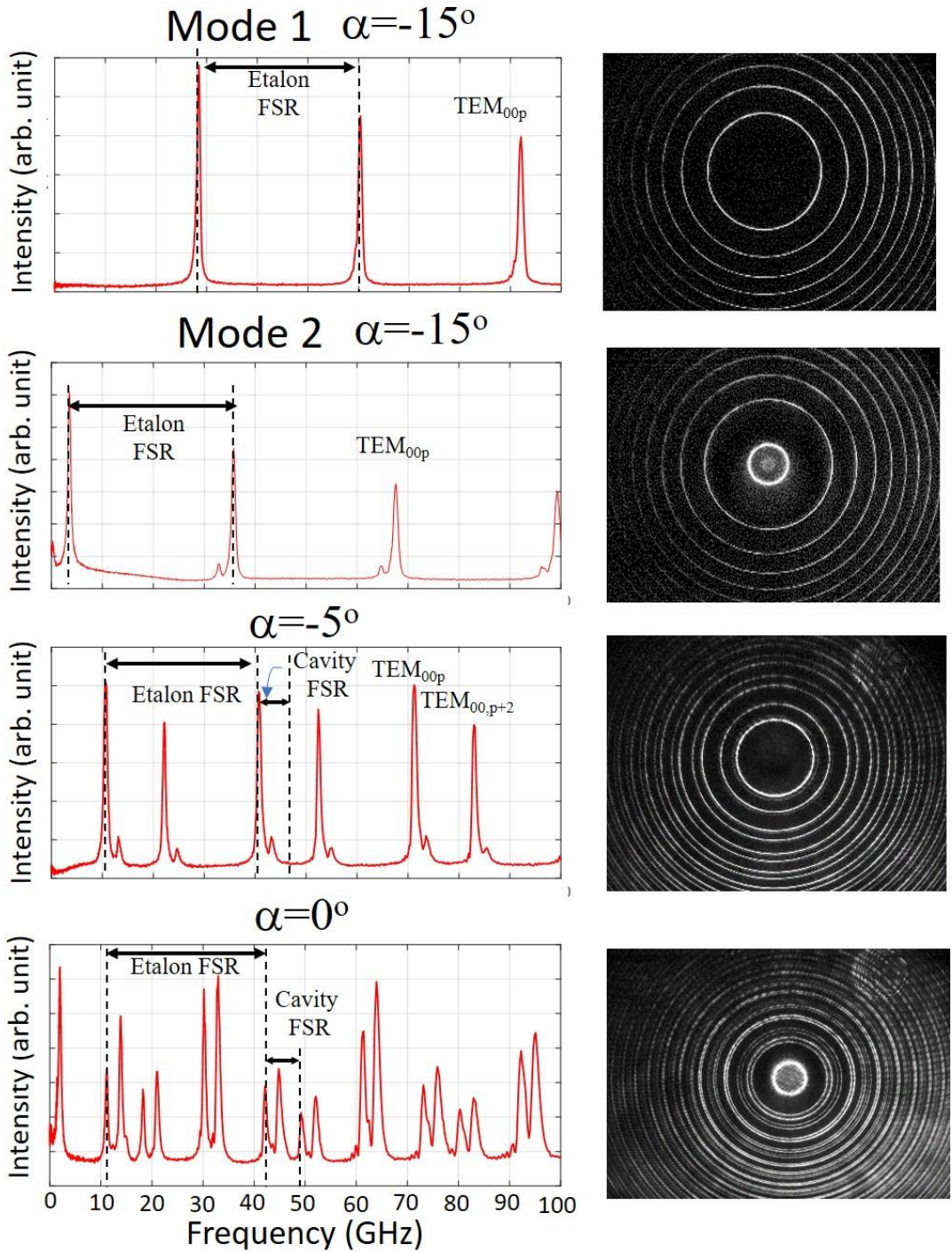

Fig. 4 Frequency content of the emitted beam for the broken PT-symmetric at α= -15° (top two panels) for both polarization eigenstates, near the EP at α=-5° (third panel) and in the unbroken symmetric region at α=0° (lower panel). The corresponding interference patterns are shown on the right.

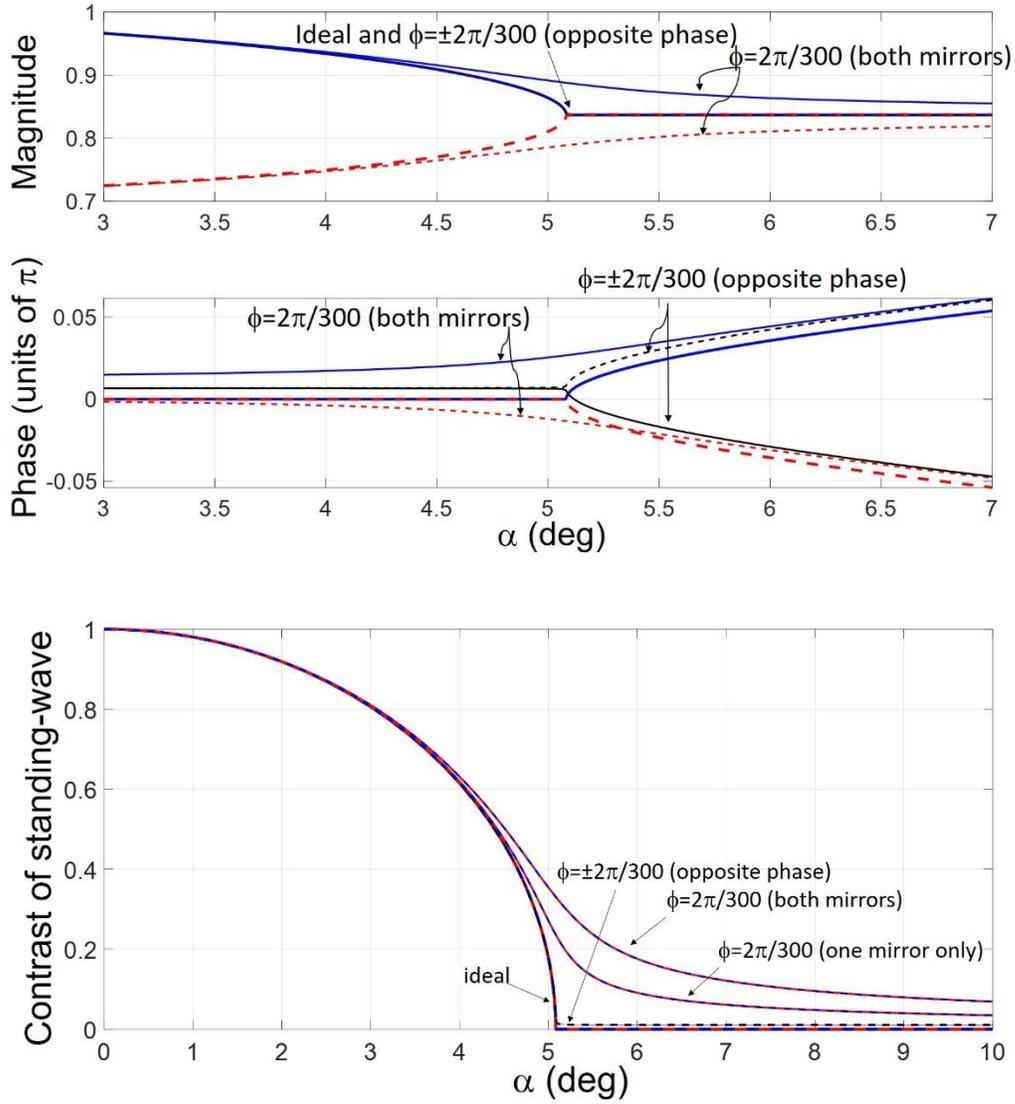

Fig. 5. (a) Effect of a small error on the mirror phase shift between orthogonal directions on the eigenvalue spectrum of the roundtrip Jones matrix near the transition between unbroken and broken PT-symmetry, for the perfect case discussed in the manuscript, for an identical $2\pi/300$ phase shift and for opposite phase shift values on both mirrors. (b) Effect on the contrast of the standing wave for the same conditions as shown in Fig. 5a.